\documentclass[apj]{emulateapj}
\usepackage[alpha,z]{tmsfnts}
\def\l5{CL5}
\def\cl6{CL6}
\def\cl7{CL7}
\def\cl9{CL9}
\def\cl10{CL10}
\def\cl11{CL11}
\def\cl14{CL14}
\def\cl24{CL24}
\def\l101{CL101}
\def\cl102{CL102}
\def\cl103{CL103}
\def\cl104{CL104}
\def\cl105{CL105}
\def\cl106{CL106}
\def\cl107{CL107}
\def\cl108{CL108}

\def\Lx{L_{\rm X}}
\def\Yx{Y_{\mkern-0.2mu\rm X}}
\def\Tx{T_{\mkern-0.4mu\rm X}}
\def\Mg{M_{\rm g}}
\def\Mtot{M_{\rm tot}}
\def\M500{M_{\rm 500}}

\def\hide#1{}

\begin{document}

\submitted{The Astrophysical Journal, submitted}
\vspace{1mm}
\slugcomment{{\em The Astrophysical Journal, submitted}} 
\shortauthors{KRAVTSOV, VIKHLININ \& NAGAI}
\shorttitle{A NEW X-RAY CLUSTER MASS INDICATOR}

\title{A New Robust Low-Scatter X-ray Mass Indicator for Clusters of Galaxies}

\author{Andrey V. Kra\kern-0.05emvtsov\altaffilmark{1,2}, Alexey Vikhlinin\altaffilmark{3,4},
Daisuke Nagai\altaffilmark{5} }  
\altaffiltext{1}{Department of Astronomy \& Astrophysics, The
  University of Chicago, Chicago, IL 60637 USA ({\tt andrey@oddjob.uchicago.edu})} 
\altaffiltext{2}{Kavli Institute for Cosmological Physics and Enrico
  Fermi Institute, The University of Chicago, Chicago, IL 60637 USA } 
\altaffiltext{3}{Harvard-Smithsonian Center for Astrophysics, 60 Garden Street, Cambridge, MA 02138}
\altaffiltext{4}{Space Research Institute, 8432 Profsojuznaya St., GSP-7, Moscow 117997, Russia}
\altaffiltext{5}{Theoretical Astrophysics, California Institute of Technology, Mail Code 130-33, 
Pasadena, CA 91125}

\begin{abstract}
  We present comparison of X-ray proxies for the total cluster mass,
  including the spectral temperature ($\Tx$), gas
  mass measured within $r_{500}$ ($\Mg$), and the new proxy, $\Yx$,
  which is a simple product of $\Tx$ and $\Mg$ and is related to the total
  thermal energy of the ICM. We use mock \emph{Chandra} images
  constructed for a sample of clusters simulated with the eulerian
  $N$-body$+$gasdynamics adaptive mesh refinement ART code in the
  concordance $\Lambda$CDM cosmology.  The simulations achieve high
  spatial and mass resolution and include radiative cooling, star
  formation, and other processes accompanying galaxy formation. Our
  analysis shows that simulated clusters exhibit a high degree of
  regularity and tight correlations between the considered observables
  and total mass. The normalizations of the $M-\Tx$, $\Mg-\Tx$, and
  $M-\Yx$ relations agree to better than $\approx 10-15\%$ with the
  current observational measurements of these relations. Our results
  show that $\Yx$ is the best mass proxy with a remarkably low scatter of only $\approx 5-7\%$ in
  $\M500$ for a fixed $\Yx$, at both low and high redshifts and
  regardless of whether clusters are relaxed or not. In addition, we
  show that redshift evolution of the $\Yx-M_{500}$ relation is close to
  the self-similar prediction, which makes $\Yx$ a very attractive mass
  indicator for measurements of the cluster mass function from
  X-ray selected samples.
\end{abstract}


\section{Introduction}
\label{sec:intro}

The evolution of the cluster abundance is one of the most sensitive
probes of cosmology, which can constrain the power spectrum
normalization, matter content, and the equation of state of the dark
energy. The potential and importance of these constraints have motivated
efforts to construct several large surveys of high-redshift clusters
during the next several years. However, in order to realize the full
statistical power of the upcoming cluster surveys, it is paramount that
the relation between cluster mass and observables and any potential
biases are well known.

Several cluster observables based on the galaxy velocities, optical
light, X-ray observables such as luminosity, temperature, mass of the
intracluster medium (ICM), and Sunyaev-Zel'dovich (SZ) flux have been
proposed and used in the literature as proxies of the total cluster mass
\citep[see][for a recent comprehensive review]{voit05}. In this study we
focus on the mass indicators derived from cluster X-ray observables,
which provide a handle on the properties of the hot ICM component. X-ray
luminosity, $\Lx$, computed using the flux integrated within a certain
radius or a range of radii, is expected to correlate with cluster mass
\citep[e.g.,][]{kaiser86} and is the most straightforward mass indicator
to measure observationally. $\Lx$ has been used for cosmological fits to
the cluster samples from \emph{ROSAT} All-Sky Survey
\citep{reiprich_boehringer02,allen_etal03} and Deep Cluster Survey
\citep{borgani_etal01}. However, $\Lx$ is also the least accurate
(internally) of all proposed X-ray proxies for $\Mtot$. $\Lx$ is
dominated by the cluster cores and thus is particularly susceptible to
non-gravitational processes in the ICM. Given the large scatter in the
$\Lx-\Tx$ relation
\citep[e.g.,][]{david_etal93,markevitch98,ikebe_etal02}, the $\Lx-M$ relation for real clusters probably also has significant scatter
\citep{stanek_etal06}. The slope of the $\Lx-M$ relation deviates from
the self-similar prediction \citep[e.g.,][]{allen_etal03}. In addition,
X-ray luminosity is notoriously difficult to reliably model in
cosmological simulations \citep[e.g.,][]{anninos_norman96,lewis_etal00}, a
significant disadvantage given that simulations are often used to get a
handle on the expected evolution of the mass~vs.~proxy relations. These
problems could potentially be alleviated with sufficient angular
resolution by excising the flux from cluster cores, responsible for most
of the scatter \citep{markevitch98}.

The most common choice of mass proxy used to measure the cluster number
density and constrain cosmological parameters is the X-ray temperature
of the intracluster plasma
\citep[e.g.,][]{henry_arnaud91,oukbir_blanchard92,markevitch98,henry00,seljak02,ikebe_etal02,pierpaoli_etal03}.
Until recently, there was a large apparent systematic uncertainty in the
normalization of the $M-\Tx$ relation, as evidenced, for example, by a
$\approx 30-50\%$ discrepancy between observational measurements and
cosmological simulations
\citep[e.g.,][]{finoguenov_etal01,pierpaoli_etal03}.  Over the last
several years the $M-\Tx$ normalization was revised both in simulations
and observations due to (1) inclusion of more realistic physics in
cosmological simulations (e.g., radiative cooling and star formation,
\citeauthor{dave_etal02} \citeyear{dave_etal02},
\citeauthor{muanwong_etal02} \citeyear{muanwong_etal02}), (2) improved
analyses of observed clusters using more realistic gas density profiles
\citep[e.g.,][]{borgani_etal04,vikhlinin_etal06}, (3) more reliable
measurements of the cluster temperature profiles
\citep{markevitch_etal98,nevalainen_etal00,arnaud_etal05,vikhlinin_etal06},
and (4) better understanding of the meaning of the mean spectral X-ray
temperature, $\Tx$, and the use of uniform definition of $\Tx$ in
observations and in simulations analyses
\citep{mazzotta_etal04,rasia_etal05,vikhlinin06}.  The current agreement
between models and observations is $\approx 10\%$ (see below).

The scatter in the $M-\Tx$ relation is significantly reduced compared to
that in the $\Lx-M$ relation \citep[the upper limit from observations is
$\approx15\%$ in $M$ for fixed $T$ for relaxed
clusters;][]{vikhlinin_etal06}. In general, existence of a tight
relation such as $M-\Tx$ indicates that clusters are remarkably regular
population of objects with their global properties tightly related to
total mass \citep[e.g.,][]{mohr_etal99}, and scatter caused by secondary
effects such as substructure in the ICM, non-gravitational processes,
and mergers \citep{ohara_etal06}. 

More recently, gas mass was used as a proxy for $\Mtot$
\citep{vikhlinin_etal03,voevodkin_vikhlinin04}. The practical advantage
of gas mass over temperature is that it can be measured robustly from
the X-ray imaging alone. Also, the $M_g-M$ relation in principle permits
external calibration from the CMB measurements of the global
baryon-to-dark matter ratio. Finally, $M_g$ can be expected to be less
sensitive to mergers which should translate into smaller scatter in the
$M_{\rm g}-M$ relation. The caveat is that trend of gas mass with
cluster mass and evolution with redshift are not yet fully understood. 

The use of clusters as efficient probes for precision cosmology puts
stringent requirements on observable cluster mass proxies: 1) tight,
low-scatter correlation between the proxy and mass, 2) with the
scatter insensitive to mergers, as the frequency of mergers is
expected to increase sharply with redshift \citep*{gottloeber_etal01},
3) simple power-law relation and evolution which can be described by a
small number of parameters and be as close as possible to the
prediction of the self-similar model.  The point (3) is crucial to
ensure that the self-calibration strategies for analyses of large
cluster surveys
\citep{levine_etal02,hu03,majumdar_mohr03,majumdar_mohr04,lima_hu04,lima_hu05,wang_etal04}
are successful. It is particularly important that the scatter in the
observable-mass relation is low and well-behaved \citep{lima_hu05}. 

In general, a mass proxy does not have to be a single cluster
property, such as $\Lx$, $\Tx$ or $\Mg$. Any physically-motivated
combination of these variables that is expected to be tightly related
to cluster mass can be used to construct a valid mass indicator. A
hint for a better X-ray mass proxy is provided by recent studies based
on cosmological simulations of cluster formation
\citep{motl_etal05,nagai06}, which show that integrated SZ flux,
$Y_{\rm SZ}$, proportional to the product of gas mass and temperature,
is a good, robust mass indicator with low scatter in the $Y_{\rm
SZ}-M$ relation, regardless of the dynamical state of the cluster.  In
addition, the $Y_{\rm SZ}-M$ relation exhibits a simple, nearly
self-similar evolution with redshift
\citep{dasilva_etal04,nagai06}. The physical reason for the robustness
of the SZ flux is straightforward: $Y_{\rm SZ}$ is directly related to
the total thermal energy of the ICM and thus to the depth of the
cluster potential well (see eq.~\ref{eq:ysz} in \S~\ref{s:indicators}
below).

In this study, we show that a similar robust, low-scatter mass
indicator can be constructed using X-ray observables. The indicator,
which is simply the product of the total ICM mass and X-ray
spectroscopic temperature, $Y_{\rm X}=M_{\rm g}\Tx$, correlates
strongly with cluster mass with only $\approx 5-8\%$ intrinsic
scatter. The scatter is robust to mergers, in the sense that even for
disturbed unrelaxed systems it gives unbiased estimates of mass with
the statistical uncertainty similar to that for relaxed systems. Thus,
the scatter of the $Y_{\rm X}-M$ relation at higher redshift is
similar to the scatter at $z=0$.  In addition, we show that evolution
of the slope and normalization of the $Y_{\rm X}-M$ relation is nearly
self-similar.  These properties make $Y_{\rm X}$ particularly useful
for measurements of cluster mass function using X-ray surveys.

\section{Mass Proxies}
\label{s:indicators}

Physical properties of virialized systems, such as clusters, are
expected to correlate with their total mass. For example, in the
self-similar model \citep{kaiser86,kaiser91} the cluster gas mass is
expected to be simply proportional to the total mass:
\begin{equation}\label{eq:mg-m:self}
M_{\Delta_c}=C_{M_g} M_{\rm gas}, 
\end{equation}
where masses are determined within a radius enclosing
a certain overdensity $\Delta_c$ with respect to the critical
density of the universe at the epoch of observation, $\rho_{\rm crit}(z)$, 
and $C_{M_g}$ is a constant independent of cluster mass
and redshift. The self-similar relation between cluster mass and 
temperature is:
\begin{equation}\label{eq:m-t:self}
E(z)\,M_{\Delta_c}=C_T T^{3/2}. 
\end{equation}
Here the function $E(z)\equiv H(z)/H_0$ for a flat cosmology with the
cosmological constant assumed throughout this study is given by
\citep[e.g.,][]{peebles93}:
\begin{equation}
E(z)=\sqrt{\Omega_M(1+z)^3 + \Omega_{\Lambda}},
\end{equation}
where $\Omega_M$ and $\Omega_{\Lambda}$ are the present-day
density parameters for matter and cosmological constant. 

The SZ flux integrated within a certain radius, $Y_{\rm SZ}$, is
proportional to the total thermal energy of the ICM gas and thus to the
overall cluster potential, which makes it relatively insensitive to the
details of the ICM physics and merging:
\begin{equation}\label{eq:ysz}
Y_{\rm SZ}=\left(\frac{k_B\,
\sigma_T}{m_e c^2}\right) \int_V n_e T_e\, dV\propto M_g T_m,
\end{equation}
where $k_B$, $\sigma_T$, $m_e$, and $c$ have their usual meaning, $n_e$
are $T_e$ the electron number density and temperature of the gas, and
$T_m$ is the gas mass-weighted mean temperature of the ICM. 
Combinations of
equations~(\ref{eq:mg-m:self},\ref{eq:m-t:self},\ref{eq:ysz}) gives
self-similar prediction for the $Y_{\rm SZ}-M$ relation,
\begin{equation}
\label{eq:yszm}
E(z)^{2/5}M_{\Delta_c}= C_{Y_{\rm SZ}}  Y_{\rm SZ}^{3/5}
\end{equation}

Cosmological simulations show that $Y_{\rm SZ}$ is a good, low-scatter
cluster mass proxy and that $Y_{\rm SZ}-M$ relation form and evolution
are close to the self-similar prediction
\citep{dasilva_etal04,motl_etal05,hallman_etal06,nagai06}. Given the
good qualities of $Y_{\rm SZ}$ as a mass proxy, it is interesting
whether a similar indicator can be constructed from the X-ray
observables, which could be used in studies of the X-ray cluster
abundances. The simplest X-ray analog of $Y_{\rm SZ}$ is
\begin{equation}\label{eq:Yx:def}
\Yx = \Mg\, \Tx,
\end{equation}
where $\Mg$ is the gas mass derived from the X-ray imaging data (it is
measured within a radius enclosing overdensity $\Delta_c$), and $\Tx$ is
the mean X-ray \emph{spectral} temperature
\citep{mazzotta_etal04,vikhlinin06}. As we describe below in
\S~\ref{sec:mock}, it is advantageous to measure $\Tx$ excluding the
central cluster region, which can be achieved with moderate angular
resolution X-ray telescopes ($\lesssim15''$ FWHM). To excise the central
regions is desirable because the observed cluster temperature profiles
show a greater degree of similarity outside the core
\citep{vikhlinin_etal06}, and also because this makes the spectral
temperature closer to the gas mass averaged $T_m$ which ideally should
be used in equation~\ref{eq:Yx:def}.

%

\def\P0{\phantom{,}}
\begin{deluxetable}{p{2.5cm}ccl}
\tablecaption{Simulated cluster sample at $z=0$\label{tab:sim}
}
\tablehead{
\multicolumn{1}{c}{Cluster}&
\multicolumn{1}{c}{$M^{\rm tot}_{\rm 500}$} &
\multicolumn{1}{c}{$M^{\rm gas}_{500}$} &
\multicolumn{1}{c}{$\langle \Tx\rangle$} 
\\[3pt]
\multicolumn{1}{c}{ID}&
\multicolumn{2}{c}{($10^{13}h^{-1}\rm\ M_{\odot}$)}&
\multicolumn{1}{c}{(keV)} 
}
\startdata
CL101 \dotfill & 90.8 & 8.17  &\P0 8.7 \\
CL102 \dotfill & 54.5 & 4.82  &\P0 5.8 \\
CL103 \dotfill & 57.1 & 4.91  &\P0 4.8 \\
CL104 \dotfill & 53.9 & 5.15  &\P0 7.7 \\
CL105 \dotfill & 48.6 & 4.71  &\P0 6.2 \\
CL106 \dotfill & 34.7 & 3.17  &\P0 4.3 \\
CL107 \dotfill & 25.7 & 2.17  &\P0 3.9 \\
CL3   \dotfill & 20.9 & 1.91  &\P0 3.6 \\
CL5   \dotfill & 13.1 & 1.06  &\P0 2.4 \\
CL6   \dotfill & 16.8 & 1.38  &\P0 3.4 \\
CL7   \dotfill & 14.1 & 1.21  &\P0 2.9 \\
CL9   \dotfill & 8.23 & 0.73  &\P0 1.6 \\
CL10  \dotfill & 6.72 & 0.43  &\P0 1.9 \\
CL11  \dotfill & 8.99 & 0.78  &\P0 2.0 \\
CL14  \dotfill & 7.69 & 0.62  &\P0 1.8 \\
CL24  \dotfill & 3.47 & 0.26  &\P0 1.0 
\enddata
\end{deluxetable}

\section{Mock \emph{Chandra} Images and Analyses\\ of Simulated Clusters
}
\label{sec:mock}

A detailed account of the simulations, mock image generation, and analysis 
will be presented elsewhere (Nagai, Vikhlinin \& Kravtsov 2006, 
in preparation). Here, we give a brief overview of the procedure
and define how the observables used in this study are derived.  

\subsection{Simulated Cluster Sample}
\label{sec:sim}

In this study, we use high-resolution cosmological simulations of
sixteen cluster-sized systems in the ``concordance'' flat {$\Lambda$}CDM
model: $\Omega_{\rm m}=1-\Omega_{\Lambda}=0.3$, $\Omega_{\rm
b}=0.04286$, $h=0.7$ and $\sigma_8=0.9$, where the Hubble constant is
defined as $100h{\ \rm km\ s^{-1}\ Mpc^{-1}}$, and $\sigma_8$ is the
power spectrum normalization on $8h^{-1}$~Mpc scale.  The simulations
were done with the Adaptive Refinement Tree (ART)
$N$-body$+$gasdynamics code \citep{kravtsov99, kravtsov_etal02}, a
Eulerian code that uses adaptive refinement in space and time, and
(non-adaptive) refinement in mass to reach the
high dynamic range required to resolve cores of halos formed in
self-consistent cosmological simulations. 

To set up initial conditions we first ran low resolution simulations of
two $80\,h^{-1}$~Mpc boxes and seven $120\,h^{-1}$~Mpc boxes, from which
we selected sixteen clusters with the virial masses ranging from $M_{\rm
  vir} \approx 7\times10^{13}$ to $2\times 10^{15}\,h^{-1}\,M_{\odot}$
for re-simulation at higher resolution. High-resolution simulations were
run using 128$^3$ uniform grid and 8 levels of mesh refinement in the
computational boxes of $120\,h^{-1}$~Mpc for CL101--107 and
$80\,h^{-1}$~Mpc for CL3--24.  These simulations achieve the dynamic
range of $128\times 2^8=32768$ and peak formal resolution of $\approx
3.66\,h^{-1}$~kpc and $2.44\,h^{-1}$~kpc, corresponding to the actual
resolution of $\approx 7\,h^{-1}$~kpc $5\,h^{-1}$~kpc for 120 and
$80\,h^{-1}$~Mpc boxes, respectively. Only the region of $\sim
(3-10)\,h^{-1}$~Mpc around the cluster was adaptively refined, the rest
of the volume was followed on the uniform $128^3$ grid. The mass
resolution corresponds to the effective $512^3$ particles in the entire
box, or the Nyquist wavelength $\lambda_{\rm Ny}=0.469\,h^{-1}$ and
$0.312\,h^{-1}$ comoving megaparsec for CL101--107 and CL3--24,
respectively, or $0.018\,h^{-1}$ and $0.006\,h^{-1}$~Mpc in the physical
units at the initial redshift of the simulations. The dark matter
particle mass in the region around the cluster was $9.1\times
10^{8}\,h^{-1}\,M_{\odot}$ for CL101--107 and $2.7\times
10^{8}\,h^{-1}\,M_{\odot}$ for CL3--24, while other regions were
simulated with lower mass resolution.

The cluster simulations used in this analysis include dissipationless
dynamics of dark matter, gasdynamics, star formation, metal enrichment
and thermal feedback due to the supernovae type II and type Ia,
self-consistent advection of metals, metallicity dependent radiative
cooling and UV heating due to cosmological ionizing background
\citep{haardt_madau96}. The cooling and heating rates take into account
Compton heating and cooling of plasma, UV heating, atomic and molecular
cooling and are tabulated for the temperature range $10^2<T<10^9$~K and
a grid of metallicities, and UV intensities using the \textsc{cloudy}
code \citep[ver. 96b4,][]{ferland_etal98}. The \textsc{cloudy} cooling
and heating rates take into account metallicity of the gas, which is
calculated self-consistently in the simulation, so that the local
cooling rates depend on the local metallicity of the gas. 

Star formation in these simulations was implemented using the
observationally-motivated recipe \citep[e.g.,][]{kennicutt98}:
$\dot{\rho}_{\ast}=\rho_{\rm gas}^{1.5}\,t_{\ast}^{-1}$, with
$t_{\ast}=4\times 10^9$~yrs. Stars are allowed to form in regions with
temperature $T<2\times10^4\,$K and gas density $n > 0.1$~cm$^{-3}$.  No
other criteria (like the collapse condition $\nabla\cdot \mathbf{v} <
0$) are used. Comparison of the runs with different choices of the
threshold for star formation, $n=10$, $1$, $0.1$, and $0.01$~cm$^{-3}$. 
shows that the threshold affects gas fractions at small radii, $r/r_{\rm
  vir}<0.1$, but its effect is negligible at the radii we consider in
this study. Further details about this simulated cluster sample will be
presented in Nagai, Vikhlinin \& Kravtsov (2006, in preparation). 

The properties of the simulated cluster sample are summarized in
Table~\ref{tab:sim}. We list the total and hot ICM masses defined within
a radius $r_{500}$ enclosing the overdensity of 500 with respect to the
critical density at $z=0$. We also list the mean spectral temperature
measured using mock \emph{Chandra} spectra from the radial region of
$(0.15-1)\,r_{500}$, as described in \S~\ref{sec:mockanalysis}. Although
the simulated cluster sample is small, the objects cover more than an
order of magnitude in mass and are simulated with very high resolution
which allows us to take into account effects of galaxy formation on the
ICM. 

In what follows, we use cluster total mass and observables measured
within $r_{500}\approx 0.5-0.6\,r_{\rm vir}$, where $r_{\rm vir}$ is a
traditional definition of cluster virial radius using the virial
overdensity $\Delta_{\rm vir}$ ($\approx 337$ at $z=0$ for the cosmology
adopted in our simulations) with respect to the mean density at $z=0$. 
This choice of the outer radius is mainly motivated by the fact that
clusters are more relaxed within $r_{500}$ compared to the outer regions
\citep{evrard_etal96,lau_etal06}. Also, the analysis of real clusters is
often limited to a similar radius because of the limited field of view. 

\subsection{Mock Chandra Images}
\label{sec:mockimages}

We first create the X-ray flux maps of the simulated clusters viewed
along three orthogonal projections.  The flux map is computed by
projecting the X-ray emission arising from the computational cells
within $3\,R_{\rm vir}$ of the cluster along a given line-of-sight,
taking into account the actual gas density, temperature, and metallicity
of each cell in a simulation output.  We compute the X-ray plasma
emission, $\Lambda_{E}(T_i,Z_i,z)$, using the MEKAL code with the
relative elemental abundance from \citet{anders_etal89}, and assuming
interstellar absorption with a hydrogen column density of
$n_H=2.0\times10^{20}\,$cm$^{-2}$. This provides expected emission
spectra in the 0.1--10~keV energy range on a spatial grid $1024\times
1024$ pixels with the size $4.88\,h^{-1}$~kpc for CL101--107 and
$2.44\,h^{-1}$~kpc for CL3--24. The entire map is therefore
$5.0\,h^{-1}$~Mpc and $2.5\,h^{-1}$~Mpc, respectively. In generating the
spectral maps we assume the redshift of cluster observation is $z_{\rm
  obs}=0.06$ for the simulation output at $z=0$ and $z_{\rm obs}=0.6$
for the $z=0.6$ sample. 

Next, we simulate the photon map by convolving the spectrum from each
image pixel with the on-axis response of the \emph{Chandra} ACIS-I
camera and drawing the number of photons in each spectral channel from
the Poisson distribution. We simulate two sets of the mock
\emph{Chandra} photon maps. In the first set we assume a $100$~ksec
\emph{Chandra} exposure, which is fairly typical for the real deep
observations. From this set, we generate cluster images in the
0.7--2~keV band. We then add the uniform Poisson background with the
intensity typical for ACIS-I data \citep{markevitch_etal03}. These
images are used to identify and mask out from the further analysis all
detectable small-scale clumps, as routinely done by observers.  These
clumps contain a small fraction of gas mass and do not bias gas mass
estimates.  They can, however, bias temperature measurement
significantly. Our clump detection is fully automatic and based on the
wavelet decomposition algorithm described in \citet{vikhlinin_etal98}. 

All further analysis is performed using the second set of photon maps
generated for very long exposures, $10^6$~sec for the $z=0$ sample and
$10^8$~sec exposure for the $z=0.6$ sample. Here, the exposure time is
chosen to ensure $\approx 10^6-10^7$ photons outside the cluster core
region for all simulated clusters.  This second set of data is used in
our analysis to derive gas mass and mean spectroscopic temperature of
the ICM. The exposures are artificially large by design as we are
interested in the intrinsic scatter of the X-ray observable mass
relation, not the statistical errors due photon noise in a particular
choice of short exposure. Also, we ignore further complications present
in reduction of real \emph{Chandra} data, including background
subtraction and spatial variations of the effective area (i.e., we
assume that accurate corrections for these effects can be applied to the
real data and any associated uncertainties are included in the reported
measurement errors).

\subsection{Analysis of Mock Chandra Data}
\label{sec:mockanalysis}

We analyze the mock data for the simulated clusters employing the
techniques and tools used in analyses of the real \emph{Chandra} cluster
observations, as described in detail by \citet{vikhlinin_etal06}. After
masking out detectable substructures (see above), we fit the X-ray
spectra in concentric annuli to measure the projected temperature and
metallicity profiles. Next, we measure the X-ray surface brightness
profile in the 0.7--2~keV band (used in the real data analysis because
it provides the optimal signal-to-noise ratio). Using the effective area
as a function of energy and observed projected temperature and
metallicity at each radius, we convert the observed brightness from
units of count rate to the projected emission measure, $\mathit{EM}=\int
n_e \,n_p\,dl$. The derived projected emission measure profile is fit to
a three-dimensional model of $\rho_{\rm gas}(r)$ that has great
functional freedom and can independently describe the gas density slopes
at both $r\gtrsim r_{500}$, and at smaller radii and in the very inner
region. The best fit model directly provides the gas mass profile,
$M_{\rm gas}(r)$.

We also measure the average X-ray spectral temperature, $\Tx$ from a
single-temperature fit to the spectrum integrated within $r_{500}$,
excluding the central region strongly affected by radiative cooling. The
inner cut is set at a fixed fraction $r_{500}$, $r_{\rm
  in}=0.15\,r_{500}$. Note that we choose to cut out central region
defined using a fixed fraction of $r_{500}$ rather than a fixed metric
radius of 70~kpc as in \citet{vikhlinin_etal06}. This new definition for
$\Tx$ results in only a small correction ($-3\%$ on average) of the
$\Tx$ values reported in \citet[][who used notation of $T_{\rm spec}$
instead of $\Tx$]{vikhlinin_etal06}. We choose the specific value of
$r_{\rm in}=0.15\,r_{500}$ because beyond this radius the observed
temperature profiles of clusters are approximately self-similar, while
at smaller radii the profiles show a large scatter. The choice of
$r_{\rm in}$ thus should maximize the self-similarity of the relation
between mass and $\Tx$. 

In our analysis below we distinguish the unrelaxed clusters from the
relaxed systems to test for sensitivity of the mass proxies to mergers
and substructure.  Specifically, as is usually done to classify observed
clusters, we visually examine mock $100$~ksec \emph{Chandra} images for
$x$, $y$, and $z$ projections of each cluster and classify as {\em
  ``relaxed"} clusters which have regular X-ray morphology and no
secondary maxima and minimal deviations from the elliptical symmetry. 
{\em ``Unrelaxed''} clusters are those with secondary maxima,
filamentary X-ray structures, or significant isophotal centroid shifts.

\section{Comparison of Mass Indicators}
\label{s:compind}

In this section we compare three indicators of cluster mass: X-ray
spectral temperature $\Tx$, gas mass $\Mg$, integrated SZ flux $Y_{\rm
  SZ}$, and its X-ray analog $\Yx\equiv \Mg\,\Tx$. 

\begin{figure}[tb]
\includegraphics[width=0.99\linewidth,bb=18 177 547 660,clip]{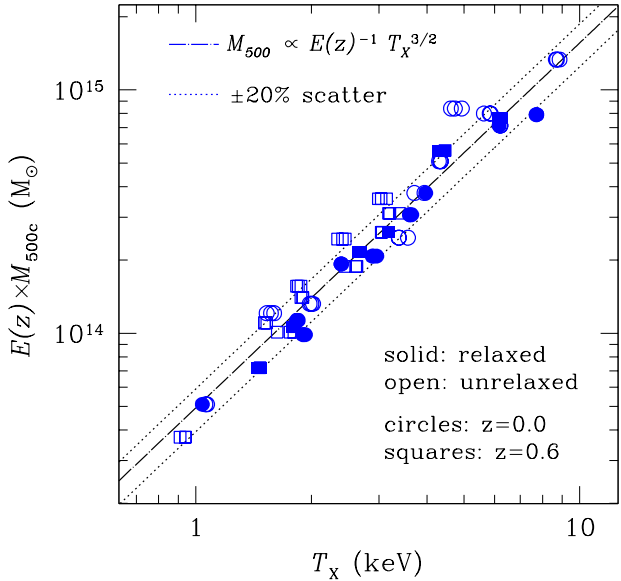}
\vspace*{0.25\baselineskip}
\caption{Relation between the X-ray spectral temperature, $\Tx$, and
  total mass, $M_{500}$. $\Tx$ is measured within the radial range
  $[0.15-1]r_{500}$. Separate symbols indicate relaxed and unrelaxed
  clusters, and also $z=0$ and $z=0.6$ samples. The dashed line shows
  the power law relation with the self-similar slope fit to the entire
  sample, and the dotted lines indicate $20\%$ scatter. }
\label{fig:txm}
\end{figure}

\begin{figure}[tb]
\includegraphics[width=0.99\linewidth,bb=18 177 547 660,clip]{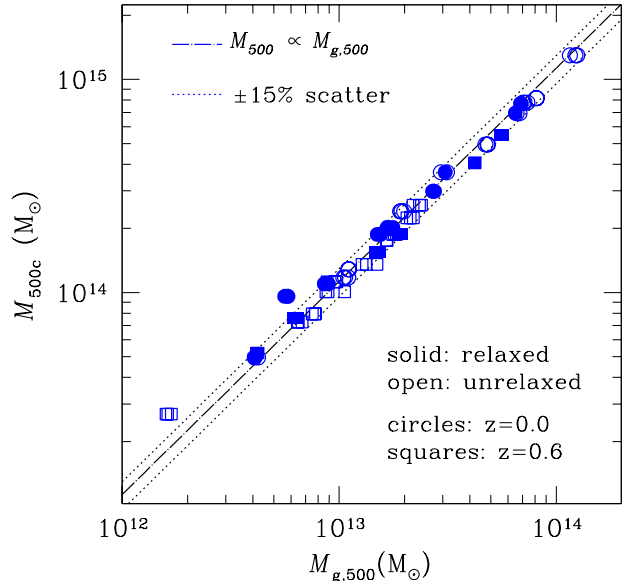}
\vspace*{0.25\baselineskip}
\caption{Correlation between gas mass and total mass of the
clusters. Both masses are measured within $r_{500}$. The meaning of
the symbols and lines is the same as in Fig.~\ref{fig:txm}. The dotted
lines indicate $15\%$ scatter.  }
\label{fig:mgm}
\end{figure}

\begin{figure}[tb]
\includegraphics[width=0.99\linewidth,bb=18 177 547 680,clip]{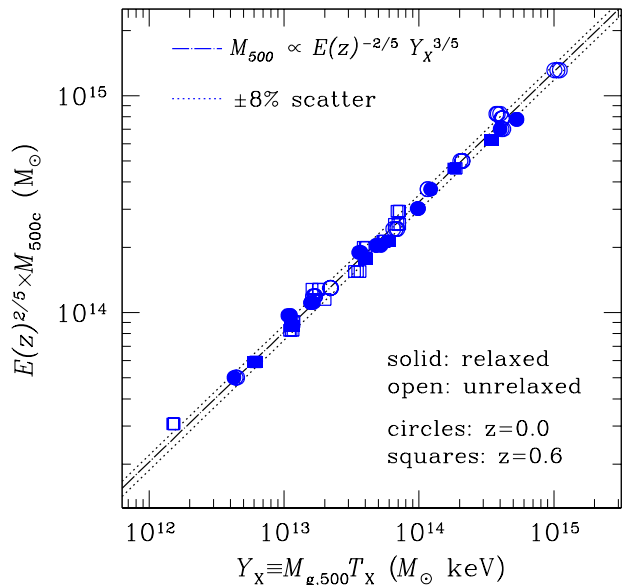}
\vspace*{0.25\baselineskip}
\caption{$\Yx-M_{500}$ correlation. The meaning
of the symbols and lines is the same as in Fig.~\ref{fig:txm}. The 
dotted lines indicate $8\%$ scatter. } 
\label{fig:yxm}
\end{figure}

Figures~\ref{fig:txm}--\ref{fig:yxm} show the $\M500-\Mg$, $\M500-\Tx$,
and $\M500-\Yx$ relations, respectively. We present correlations at
redshifts $z=0.0$ and $z=0.6$, classifying clusters into relaxed and
unrelaxed based on their X-ray image morphology, as described in
\S~\ref{sec:mockanalysis}. We performed power law fits to these
relations and present best fit values of parameters, as well as the
amount of scatter around the best fit relation for different subsets of
clusters in Table~\ref{tab:plfit}. The upper half of the table shows
fits for best fit power law normalization and scatter for the values of
slope fixed to the value predicted by self-similar model.  The lower
half shows best fit values for the case when both normalization and
slopes were allowed to vary. 

For comparison, we also include fits to the $Y_{\rm SZ}-M$ relation,
where $Y_{\rm SZ}$ is the three-dimensional integrated SZ flux
measured within $r_{500}$ \citep{nagai06}. The power law fits were
performed for each projection (i.e., $x$, $y$, and $z$) separately. 
We estimate the uncertainties in the best fit parameters by generating
$10000$ bootstrap samples \citep{numerical_recipes} and calculating
the dispersion of the best fit power law normalization and slope among
the samples. 

Figure~\ref{fig:txm} and Table~\ref{tab:plfit} show that the slope and
evolution of the $\M500-\Tx$ relation are quite close to the
self-similar model. There is a $\sim 20\%$ scatter in $M$ around the
mean relation and much of the scatter is due to unrelaxed clusters. Note
also that the normalization of the $\M500-\Tx$ relation for unrelaxed
systems is somewhat biased with respect to that for the unrelaxed
subsample. Unrelaxed clusters have somewhat lower temperatures for a
given mass. This may seem counter-intuitive at first, given that one can
expect that shocks can boost the ICM temperature during mergers. 
However, in practice the effect of shocks is relatively small
\citep[e.g.,][]{ohara_etal06}. The main source of the bias is that
during advanced mergers the mass of the system already increased but
only a fraction of the kinetic energy of merging systems is converted
into the thermal energy of the ICM \citep[see,
e.g.,][]{mathiesen_evrard01}. 

The $\Mg-\M500$ relation (Fig.~\ref{fig:mgm}) has a somewhat smaller
scatter ($\approx 10-12\%$) around the best fit power law relation
than the $M-\Tx$, but its slope is significantly different from the
self-similar prediction --- we find $\M500\propto \Mg^{0.88\div 0.92}$
compared to the expected $\M500\propto \Mg$. This is due to the trend
of gas fraction with cluster mass, $f_{\rm gas}\equiv \Mg/\M500\propto
\M500^{0.1\div 0.2}$ present for both the simulated clusters in our
sample \citep[see][]{kravtsov_etal05} and for the observed clusters
\citep{vikhlinin_etal06}.  The normalization of the $\Mg-\M500$
relation, on the other hand, evolves only weakly between $z=0.6$ and
$z=0$ (yet, statistically significant evolution is present; see
Table~\ref{tab:plfit}), reflecting slow evolution of the gas fraction
with time \citep{kravtsov_etal05}.

The $\Yx-\M500$ relation (Fig.~\ref{fig:yxm}) has the smallest scatter,
only $\approx 5-7\%$. Note that this value of scatter includes clusters at
both low and high-redshifts and both relaxed and unrelaxed systems.  In
fact the scatter in $\Yx-\M500$ for relaxed and unrelaxed systems is
indistinguishable within the errors (see Table~\ref{tab:plfit}).  Note
also that the figures include points corresponding to the three
projections of each cluster. Figure~\ref{fig:yxm} shows that the
dispersion in the projected values of $\Yx$ for each given cluster is
very small, which means that $\Yx$ is not very sensitive to the
asphericity of clusters.  Remarkably, the scatter of $\Yx-\M500$ is as
small as that in the three-dimensional $Y_{\rm SZ}-\M500$ relation (see
Table~\ref{tab:plfit}).  This implies that $\Yx$ is in fact a better
mass indicator than SZ flux, when additional possible sources of scatter
for the real measurements of the SZ flux, such as projecting structures
outside $r_{500}$, are taken into account. 

\def\nls{\\[-5pt]}
\begin{deluxetable*}{cccccc}
\tablecaption{Power-law $M_{500}=CX^{\alpha}$ best fit parameters and
  scatter\label{tab:plfit}}
\tablehead{
\nls
\multicolumn{1}{c}{sample\tablenotemark{$^a$}}&
\multicolumn{1}{c}{quantity\tablenotemark{$^b$}}&
\multicolumn{1}{c}{$M_{\rm t,500}-\Tx$} &
\multicolumn{1}{c}{$M_{\rm t,500}-M_{\rm g,500}$} &
\multicolumn{1}{c}{$M_{\rm t,500}-\Yx$} &
\multicolumn{1}{c}{$M_{\rm t,500}-Y_{\rm SZ}^{3D}$}
\nls
}
\startdata
\nls
all $z$,     & $\log_{10}C$      & $14.41\pm 0.014$ & $14.35\pm 0.010$ & $14.27\pm 0.006$ & $14.27\pm 0.006$ \\
all clusters & $\alpha$ & 1.500 & 1.000 & 0.600 & 0.600 \\
             & scatter  & 0.196 & 0.144 & 0.076 & 0.070 \\
\nls\hline\nls
all $z$,     & $\log_{10}C$     & $14.36\pm 0.016$ & $14.36\pm 0.019$ & $14.26\pm 0.009$ & $14.26\pm 0.010$ \\
relaxed      & $\alpha$ & 1.500 & 1.000 & 0.600 & 0.600 \\
             & scatter  & 0.137 & 0.157 & 0.070 & 0.080 \\
\nls\hline\nls
all $z$,     & $\log_{10}C$     & $14.44\pm 0.018$ & $14.35\pm 0.011$ & $14.28\pm 0.007$ & $14.27\pm 0.006$ \\ 
unrelaxed    & $\alpha$ & 1.500 & 1.000 & 0.600 & 0.600 \\
             & scatter  & 0.191 & 0.131 & 0.075 & 0.059 \\
\nls\hline\nls
$z=0$,       & $\log_{10}C$     & $14.39\pm 0.020$ & $14.37\pm 0.013$ & $14.27\pm 0.007$ & $14.27\pm 0.008$ \\
all clusters & $\alpha$ & 1.500 & 1.000 & 0.600 & 0.600 \\
             & scatter  & 0.221 & 0.131 & 0.068 & 0.078 \\
\nls\hline\nls
$z=0.6$,     & $\log_{10}C$     & $14.43\pm 0.017$ & $14.34\pm 0.015$ & $14.27\pm 0.009$ & $14.26\pm 0.007$ \\
all clusters & $\alpha$ & 1.500 & 1.000 & 0.600 & 0.600 \\
             & scatter  & 0.164 & 0.151 & 0.084 & 0.059 \\
\nls\hline\hline\nls
all $z$,     & $\log_{10}C$      & $14.41\pm 0.014$ & $14.35\pm 0.008$ & $14.27\pm 0.006$ & $14.27\pm 0.005$ \\
all clusters & $\alpha$ & $1.521\pm 0.062$  & $0.921\pm 0.023$ & $0.581\pm 0.009$  & $0.585\pm 0.010$ \\
             & scatter  & 0.195 & 0.107 & 0.071 & 0.067 \\
\nls\hline\nls
all $z$,     & $\log_{10}C$      & $14.36\pm 0.017$ & $14.36\pm 0.015$ & $14.26\pm 0.008$ & $14.26\pm 0.009$ \\
relaxed      & $\alpha$ & $1.533\pm 0.103$  & $0.898\pm 0.051$ & $0.579\pm 0.012$  & $0.564\pm 0.014$ \\
             & scatter  & 0.136 & 0.115 & 0.053 & 0.058 \\
\nls\hline\nls
all $z$,     & $\log_{10}C$      & $14.44\pm 0.018$ & $14.34\pm 0.010$ & $14.28\pm 0.008$ & $14.27\pm 0.006$ \\
unrelaxed    & $\alpha$ & $1.553\pm 0.063$  & $0.931\pm 0.029$ & $0.589\pm 0.010$  & $0.600\pm 0.010$ \\
             & scatter  & 0.186 & 0.095 & 0.072 & 0.059 \\
\nls\hline\nls
$z=0$,       & $\log_{10}C$      & $14.39\pm 0.019$ & $14.37\pm 0.012$ & $14.27\pm 0.007$ & $14.28\pm 0.008$ \\
all clusters & $\alpha$ & $1.524\pm 0.070$  & $0.917\pm 0.028$ & $0.583\pm 0.010$  & $0.584\pm 0.013$ \\
             & scatter  & 0.219 & 0.090 & 0.064 & 0.075 \\
\nls\hline\nls
$z=0.6$,     & $\log_{10}C$      & $14.44\pm 0.018$ & $14.31\pm 0.009$ & $14.27\pm 0.008$ & $14.26\pm 0.007$ \\
all clusters & $\alpha$ & $1.590\pm 0.086$  & $0.871\pm 0.033$ & $0.571\pm 0.016$  & $0.577\pm 0.012$ \\
             & scatter  & 0.157 & 0.077 & 0.075 & 0.051 
\nls
\enddata
\tablenotetext{$^a$}{Power law fits were performed for different
subsets of our cluster sample, split in redshift and/or dynamical
state. } \tablenotetext{$^b$}{For each observable $X$ ($=\Tx$, $\Mg$,
$\Yx$, $Y_{\rm SZ}^{3D}$), we fit power law relation of the form $\M500=C(X/X_0)^{\alpha}$, 
with $X_0=3.0$, $2\times 10^{13}\,M_{\odot}$, $4\times 10^{13}$, and $5\times 10^{-6}\, {\rm Mpc^{2}}$
for $\Tx$, $\Mg$, $\Yx$, $Y_{\rm SZ}$, respectively. }
\end{deluxetable*}

\begin{figure}[tb]
\includegraphics[width=0.99\linewidth,bb=18 160 547 665,clip]{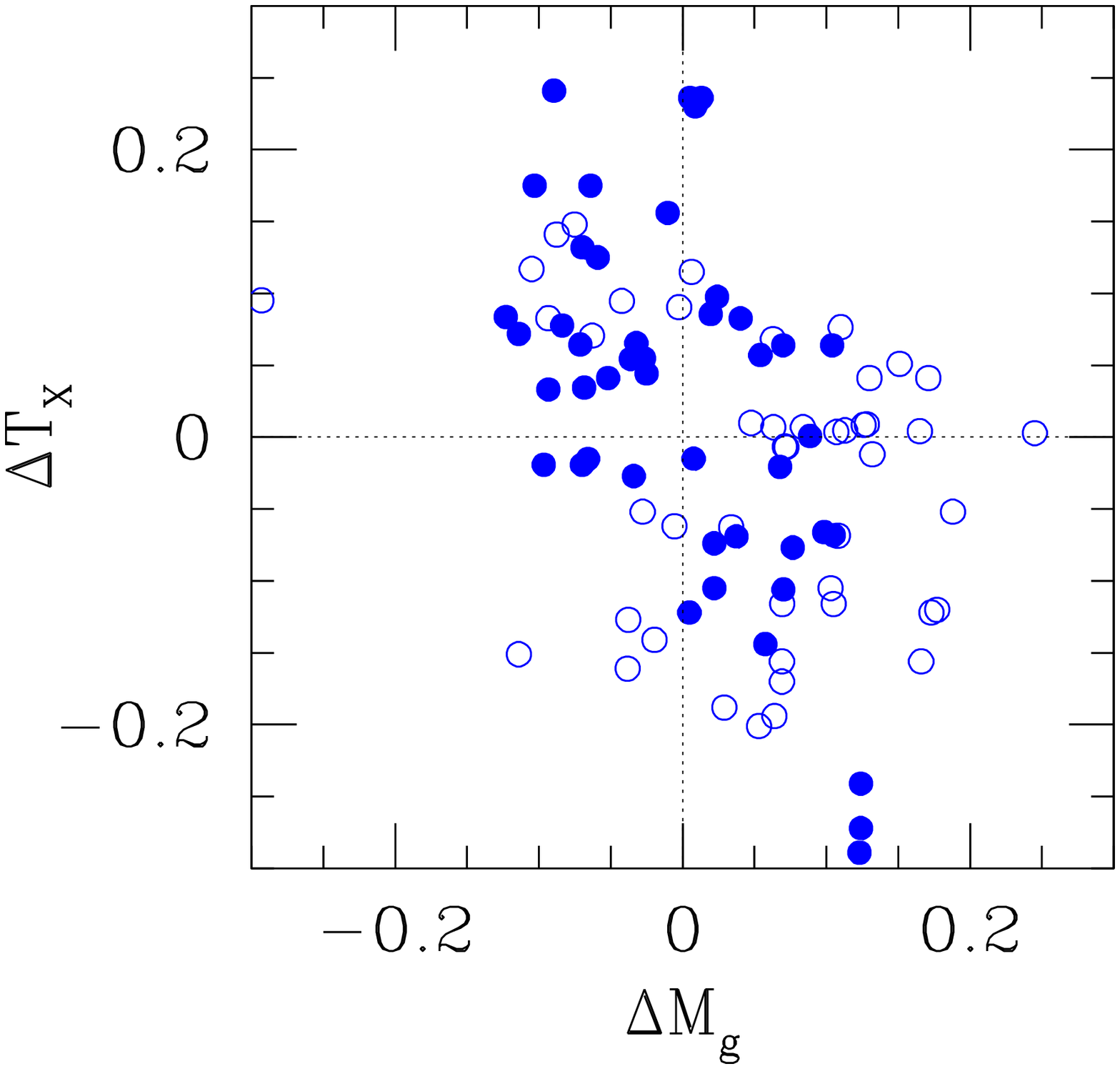}
\caption{Fractional deviations in temperature and gas mass for fixed
  $\Mtot$ relative to their respective best fit self-similar relations,
  $M_{500}\propto \Tx^{1.5}$ and $M_{500}\propto\Mg$. The fit includes
  all systems, both at $z=0$ ({\it solid circles}) and $z=0.6$ ({\it
    open circles}).  Note that the deviations for gas mass and
  temperature are generally anti-correlated: clusters with large
  positive (negative) deviations in $\Mg$ tend to have negative
  (positive) deviations in $\Tx$.  A similar anti-correlation exists in
  the trend with redshift (compare the distribution of points for $z=0$
  and $z=0.6$). }
\label{fig:res}
\end{figure}

The comparison of the mass proxies, clearly shows that $\Yx$, the
product of gas mass and X-ray spectral temperature, is by far the most
robust and most self-similar mass indicator.  Why is the product better
than its parts? Figure~\ref{fig:res} shows that the answer lies in the
anti-correlation of the residuals of temperature and gas mass from their
respective relations with total cluster mass. We plot residuals from the
best fit power laws with the slope value fixed to the self-similar value
(i.e., the upper portion of Table~\ref{tab:plfit}) to illustrate both
random scatter and systematic deviations from self-similar behavior.

Figure~\ref{fig:res} also shows that the clusters with temperatures
lower than the mean temperature for a given total mass tend to have gas
mass higher than the mean, and vice versa. Note also that there is some
redshift evolution between $z=0$ and $z=0.6$ --- more clusters have
negative deviations of temperature and positive deviations of measured
gas mass at $z=0.6$ compared to $z=0$. This redshift evolution is thus
in the opposite direction for the gas mass and temperature deviation. 
The measured $\Mg$ systematically increases at higher $z$ for a fixed
total mass, because the clusters become less relaxed on average. For
unrelaxed clusters, the ICM density distribution is non-uniform which
results in overestimation of $\Mg$ from the X-ray data
\citep{mathiesen_etal99}. Some of the decrease of $\Mg$ at lower $z$ may
be due to continuing cooling of the ICM which decreases the mass of hot,
X-ray emitting gas.

The anti-correlation of residuals and opposite evolution with redshift
for gas mass and temperature is the reason why the behavior of their
product, on average, has smaller scatter and is closer to the
self-similar expectation in both the slope and evolution. We discuss the
origin of this behavior further in \S~\ref{s:discussion} below. 

\section{Practical algorithm for estimating cluster mass using $\Yx$}
\label{sec:practical}

Suppose we have $\Yx-\M500$ relation pre-calibrated by some external
means: $\M500=C\,E(z)^{\gamma}\,\Yx^{\alpha}$, and we would like to use
it to estimate $\M500$.
The pre-calibration can be done using a well-observed sample of relaxed
clusters or simulations. Note that the definition of $\Yx$ includes
spectral temperature and gas mass within $r_{500}$. In practice,
however, when we want to use $\Yx$ to estimate $\M500$ we do not know
the value of $r_{500}$ a priori. This is not a problem for measurements
of $\Tx$, because the integrated spectral temperature is dominated by
the signal from the inner, brighter regions of the clusters and is thus
not sensitive to the exact choice of the outer radius. For example, we
checked that for clusters in the \citet{vikhlinin_etal06} sample, a
fairly drastic change of the outer radius to $0.5 r_{500}$ from the
fiducial choice of $r_{500}$ adopted in our analysis, results in a
systematic increase of the mean spectral temperature of only $\approx
5\%$.  The sensitivity to the exact choice of the inner region is also
comparably small, as was discussed above in \S~\ref{sec:mockanalysis}.

The measurement of gas mass, however, will depend quite sensitively on
the adopted outer radius, because the profile $M_g(r)$ does not converge
at large $r$.  In practice, therefore, we have to solve for $r_{500}$
simultaneously with estimating $M_{500}$, which can be done with the
following iterative algorithm. An initial value for $\Tx$ can be
obtained by using the integrated temperature (including the center) to
estimate an approximate value of $r_{500}$ through a crude
mass-temperature relation, and then re-measuring the spectrum in the
radial range $(0.15-1)\,r_{500}$. Then, using identity $M_{500} = 4/3\,\pi\,
500\, \rho_c(z)\, r_{500}^3$, we can re-write the $\Yx-M$ relation as
\begin{equation}\label{eq:Yx:Mest}
  4/3\, \pi\, 500\, \rho_c(z)\, r_{500}^3 = C\, E(z)^{\gamma}\,\left[\Yx(r_{500})\right]^{\alpha},
\end{equation}
where $\Yx(r)\equiv \Tx\,\Mg(r)$ and the constant $C$ is from
pre-calibrated $\Yx-\M500$ relation. Equation~(\ref{eq:Yx:Mest}) is used
to solve for $r_{500}$.  A second iteration can then be done using $\Tx$
re-measured in the newly estimated radial range $(0.15-1)\,r_{500}$,
continuing the procedure until convergence. Once $r_{500}$ is
determined, $M_{500} = 4/3\,\pi\,500\, \rho_c(z)\, r_{500}^3$. 

It is useful to know how the observational calibration of the
normalization constant $C$ in equation~\ref{eq:Yx:Mest} scales with the
assumed value of the Hubble constant. Cluster masses determined
dynamically or through gravitational lensing scale as $M\propto h^{-1}$. 
This implies that the angular size that corresponds to $r_{500}$ is
$h$-independent. The gas mass derived from X-ray data for a fixed
angular radius scales as $M_g\propto h^{-5/2}$.  Therefore, the observed
normalization of the $\Yx-M$ relation with the slope close to
self-similar scales as $C\propto h^{1/2}$.

\section{Discussion and Conclusions}
\label{s:discussion}

In this paper we presented comparison of several X-ray proxies for the
cluster mass --- the spectral temperature and gas mass measured within
$r_{500}$, and the new proxy, $\Yx$, defined as a simple product of
$\Tx$ and $\Mg$. Analogously to the integrated Sunyaev-Zel'dovich flux,
$\Yx$ is related to the total thermal energy of the ICM. To test these
mass proxies, we use a sample of clusters simulated in the concordance
$\Lambda$CDM cosmology.  The simulations achieve high spatial resolution
and include radiative cooling, star formation, and other processes
accompanying galaxy formation. We construct mock \emph{Chandra} images
for the simulated clusters and derive the X-ray proxies from a procedure
essentially identical to that used in the real data analysis. 

The main result of this study is that $\Yx$ is a robust mass indicator
with remarkably low scatter of only $\approx 5-7\%$ in $\M500$ for fixed
$\Yx$, regardless of whether the clusters are relaxed or not. In
addition, we show that the redshift evolution of the $\Yx-M_{500}$
relation is close to the self-similar prediction given by
equation~\ref{eq:yszm}, which makes this indicator a very attractive
observable for studies of cluster mass function with the X-ray
selected samples. 

The $\Tx-\M500$ relation has the largest scatter ($\approx 20\%$), most
of which is due to unrelaxed clusters. The unrelaxed clusters have
temperatures biased low for a given mass. This is likely because during
advanced, ongoing mergers, the mass of the system already increased but
only a fraction of the kinetic energy of merging systems is converted
into the thermal energy of gas, due to incomplete relaxation. 

The $\Mg-\M500$ relation shows an intermediate level of scatter,
$\approx 10-12\%$. This relation does not appear to be sensitive to
mergers. It does, however, exhibit significant deviations from
self-similarity in its slope, which is due to the dependence of gas
fraction within $r_{500}$ on the cluster mass \citep{kravtsov_etal05}. A
similar dependence exists for the observed clusters
\citep{vikhlinin_etal06} and we can thus expect similar trends in the
$\Mg-\M500$ relation for real clusters. 

Generally, all the observable--mass relations we tested demonstrate a
remarkable degree of regularity of galaxy clusters as a population. $\Tx$,
$\Mg$, and $\Yx$ all exhibit correlations with $\M500$ which are close to
the expectation of the self-similar model, both in their slope and
evolution with time, within the uncertainties provided by our
sample. The only exception is the slope of the $\Mg-\M500$ relation.

\begin{figure}[tb]
\includegraphics[width=0.97\linewidth,bb=18 177 547 670,clip]{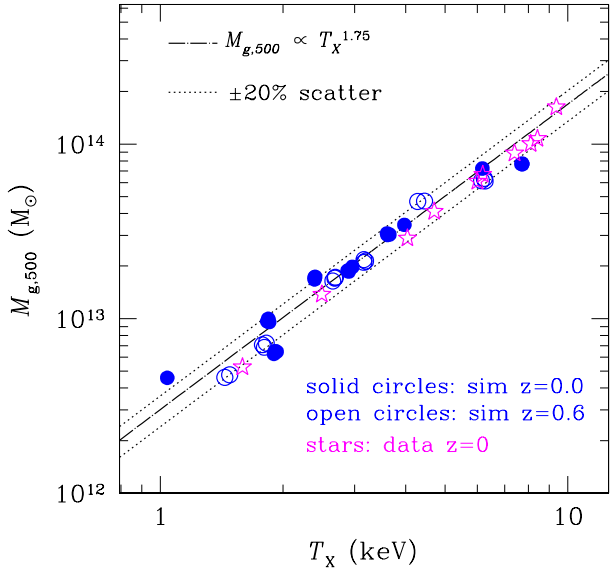}
\caption{Relation betwen X-ray spectral temperature and gas mass for the
  relaxed subsample of simulated clusters (circles) and for a sample of
  relaxed \emph{Chandra} clusters of
  \citet[][\emph{stars}]{vikhlinin_etal06}. Both gas mass and
  temperature are the quanities derived from analysis of real and mock
  X-ray data. The error bars in the \emph{Chandra} measurements are
  comparable to the symbol size and are not shown for clarity.  The gas
  masses for the simulated clusters are rescaled by a factor of
  $0.17/0.143=1.19$ to reflect the difference between the universal
  baryon fractions adopted in the simulation and the value measured by
  the \emph{WMAP}.  The \emph{dashed line} shows the best fit power law
  relation with the slope $1.75$.} 
\label{fig:tmg}
\end{figure}

Given that our analysis relies on cosmological simulations, it is
reasonable to ask whether the simulated clusters are realistic. 
Although simulations certainly do not reproduce all of the observed
properties of clusters, especially in their core regions, the ICM
properties outside the core in simulations and observations agree quite
well. We illustrate this in Fig.~\ref{fig:tmg}, which shows that the
$M_{g,500}-\Tx$ relations for simulated and observed clusters \citep[the
sample of relaxed clusters at $z\approx 0$ from][]{vikhlinin_etal06}. 
For this comparison we use only those clusters from simulations that
appear regular and relaxed in their X-ray surface brightness image. 
Clearly, both simulated and observed clusters exhibit tight correlations
between $M_{g,500}$ and $\Tx$ \citep[see
also][]{mohr_etal99,voevodkin_etal02b} which agree remarkably in their
slope ($M_g \propto T^{1.75}$) and
normalization\footnote{\label{fn:rescale}For this comparison, the gas
  masses of simulated clusters are rescaled by a factor of
  $0.17/0.143=1.19$ to reflect the difference between the universal
  baryon fractions adopted in the simulation and the value measured by
  the \emph{WMAP} \citep{spergel_etal03,tegmark_etal04}.}.  The
normalizations derived from simulated and real clusters agree to
$\approx 10\%$, while slopes are indistiguishable and both deviate
significantly from the expected self-similar value of $1.5$. This is a
consequence of significant trends in the gas fraction with cluster mass,
$M_{g,500}/M_{500}\propto M_{500}^{0.2\div 0.25}$ for both simulated
\citep{kravtsov_etal05} and observed clusters \citep{vikhlinin_etal06}. 
The deviations from the self-similar model also manifest themselves in
the absence of any noticeable evolution with redshift\footnote{Note that
  $M_g$ in Fig.\ref{fig:tmg} is not multiplied by the $E(z)$ factor
  unlike the total mass in Fig.\ref{fig:txm} and \ref{fig:yxm}.}. 
Interestingly, the real clusters show a similarly weak evolution in the
$\Mg-T$ relation \citep{vikhlinin_etal02}. Figure~\ref{fig:res} shows
that the likely explanation is that the clusters at $z=0.6$ tend to be
colder for the fixed $\Mtot$ but have higher estimated $M_{g,500}$ then
their counterparts at $z=0$ because they are less relaxed. 

A similar level of agreement between the simulations and latest
\emph{Chandra} measurements exists also for the total mass vs.{}
temperature relation, $M_{500}-\Tx$. In fact, the normalization for our
simulated sample (Table~\ref{tab:plfit}) agrees with the observational
results of \citet{vikhlinin_etal06} to $\approx 10\%$. This is a
considerable improvement given that significant disagreements existed
just several years ago (see \S~\ref{sec:intro}). The residual systematic
$10\%$ difference in the normalization is likely caused by non-thermal
pressure support from bulk gas motions
\citep{faltenbacher_etal05,rasia_etal06,lau_etal06}, which is
unaccounted for by the X-ray hydrostatic mass estimates. 

\begin{figure}[t]
\includegraphics[width=0.97\linewidth,bb=18 177 547 670]{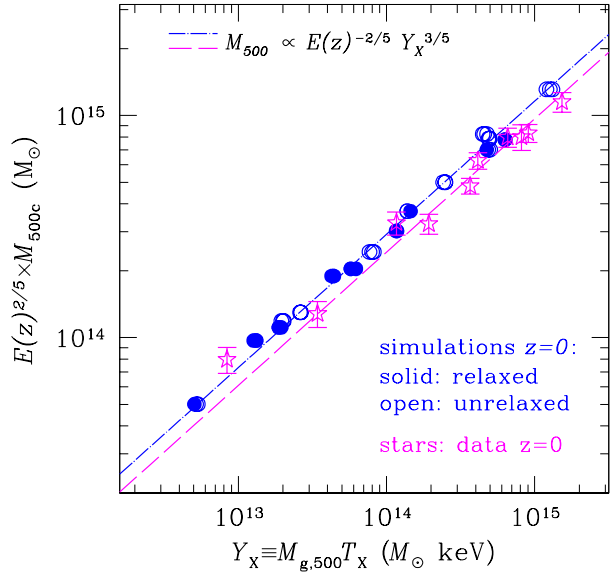}
\caption{$\Yx-\M500$ relation for the $z=0$ sample of the simulated
  clusters (\emph{circles}) and for a sample of relaxed \emph{Chandra}
  clusters of \citet[][ {\it stars}]{vikhlinin_etal06}.  The gas masses
  for the simulated clusters are appropriately rescaled (see caption to
  Fig.\ref{fig:tmg}). The {\it dot-dashed} line shows the best fit power
  law relation for the simulated clusters with the slope fixed to the
  self-similar value of $3/5$. The {\it dashed} line shows the same best
  fit power law, but with the normalization scaled down by $15\%$. }
\label{fig:yxmc}
\end{figure}

In Figure~\ref{fig:yxmc} we compare the $\Yx-\M500$ relation for the
simulated clusters\footnotemark[\ref{fn:rescale}]
and for
the \emph{Chandra} sample of \citet{vikhlinin_etal06}. The observed
clusters show a tight correlation with the slope close to the
self-similar value.  There is $\approx 15\%$ difference in
normalization, likely explained also by neglecting the turbulent
pressure support in the \emph{Chandra} hydrostatic mass estimates. 

The excellent agreement of simulations and observations in terms of the
relation between the two X-ray observables used to compute $\Yx$
($\Mg-\Tx$) and a relatively good agreement in the $\Tx-\M500$ and
$\Yx-\M500$ relations, gives us confidence that the results presented in
this paper are sufficiently realistic. One can ask whether the real
clusters show the same trend of decreasing scatter when $\Mg$ and $\Yx$
are used as mass indicators instead of $\Tx$.  Unfortunately, the
existing data cannot answer this question because the mass measurement
uncertainties for individual clusters are of order or larger than the
expected scatter for relaxed clusters. 

Our results show that $\Yx$ is clearly most robust and most self-similar
X-ray cluster mass indicator. The biases existing in mass estimates
based on $\Mg$ and $\Tx$ anti-correlate both for a given redshift and in
terms of evolutionary trends (see Figure~\ref{fig:tmg}). This explains
why their product, $\Yx$, is a better mass indicator than $\Tx$ and
$\Mg$ individually. The quality of $\Yx$ compares well to that for the
actual three-dimension integral of the ICM thermal energy (proportional
to $Y_{\rm SZ}$) in terms of its low scatter and self-similarity (see
Table~\ref{tab:plfit}).  $\Yx$ may prove to be an even better mass proxy
than $Y_{\rm SZ}$, given that we use ideal 3D measurement of the latter
while reproducing the actual data analysis for the former.  Note also
that $Y_{\rm SZ}$ is more sensitive to the outskirts of clusters,
because it involves gas mass-weighted temperature (as opposed to the
spectral temperature more sensitive to the inner regions), and thus
should be more prone to projection effects. 

Note that $\Yx$ is also an attractive mass proxy from the data analysis
point of view. First, it reduces statistical noise by combining the two
independently measured quantities, $\Mg$ and $\Tx$, into a single
quantity. Consider for example how mass estimates are affected by $\Tx$,
the parameter that is the most difficult to measure and which is most
affected by the dynamical state of cluster. A $10\%$ observational
uncertainty or a real deviation due to a merger in $\Tx$ translates into a
$\sim 15\%$ mass uncertainty through the $M-\Tx$ relation and only $6\%$
uncertainty through the $\Yx-M$ relation. $\Yx$ is also less sensitive
to any errors in the absolute calibration of the X-ray telescope. For
example, a mis-calibration of the low energy effective area typically
translates into $\Tx$ and $\Mg$ measurement errors of the opposite sign,
$\delta T/T \approx -2\,\delta \Mg/\Mg$. These errors partially cancel
in $\Yx$, $\delta Y/Y \approx 0.5\, \delta T/T$, and are further reduced
in the mass estimate: $\delta M/M = 0.6\, \delta Y/Y \approx 0.3\,
\delta T/T$.  The error in mass is larger when other proxies are used,
$\delta M/M = \delta M_g/M_g \approx 0.5\,\delta T/T$ for the $M_g-M$
relation and $\delta M/M = 1.5\,\delta T/T$ for the $M-T$ relation.

The robustness and low scatter make $\Yx$ an excellent mass indicator
for observational measurements of cluster mass function at both $z=0$
and higher redshifts. The necessary data --- an X-ray brightness
profile and a wide-beam spectrum excluding the core --- are easily
obtained with sufficiently deep observations with \emph{Chandra},
\emph{XMM-Newton}, and \emph{Suzaku} telescopes. The small scatter and
simple, nearly self-similar evolution of the $\Yx-M$ relation hold
promise for the self-calibration strategies for future large X-ray
cluster surveys.

\acknowledgements This project was supported by the National Science
Foundation (NSF) under grants No.  AST-0206216 and AST-0239759, by
NASA through grant NAG5-13274, and by the Kavli Institute for
Cosmological Physics at the University of Chicago. AV is supported by
the NASA grant NAG5-9217 and contract NAS8-39073. The cosmological
simulations used in this study were performed on the IBM RS/6000 SP4
system ({\tt copper}) at the National Center for Supercomputing
Applications (NCSA). D.N. is supported by a Sherman Fairchild
postdoctoral fellowship at Caltech.  We have made extensive use of the
NASA Astrophysics Data System and {\tt arXiv.org} preprint server. 

\bibliographystyle{apj}
\bibliography{ms}

\end{document}